\documentclass{article}
\usepackage{biblatex}
\usepackage{graphicx} 

\title{Game-theoretical approach to decentralized multi-drone conflict resolution and emergent traffic flow operations}
\author{Serge Hoogendoorn, Victor L. Knoop, \\ Hani Mahmassani, and Sascha Hoogendoorn-Lanser}
\date{August 1 2023}

\begin{document}
\maketitle

\section{Abstract}
This paper introduces decentralized control concepts for drones using differential game theory. The approach optimizes the behavior of an ego drone,  assuming the anticipated behavior of the opponent drones using a receding horizon approach. For each control instant, the scheme computes the Nash equilibrium control signal which is applied for the control period. This results in a multi-drone conflict resolution scheme that is applied to all drones considered. 

The paper discusses the approach and presents the numerical algorithm, showing several examples that illustrate the performance of the model. We examine at the behavior of the ego drone, and the resulting collective drone flow operations. The latter shows that while the approach aims to optimize the operation cost of the ego drone, the experiments provide evidence that resulting flow operations are very efficient due to the self-organization of various flow patterns. 

The presented work contributes to the state of the art in providing a generic approach to multi-drone conflict resolution with good macroscopic flow performance characteristics. The approach enables relatively straightforward inclusion of error due to sensing and communication. The approach also allows for including different risk levels (e.g., for malfunctioning of sensor and communication technology), priority rules, regulations, and higher-level control signals (e.g., routing, dynamic speed limits). 

\hfill\break%
\noindent\textit{Keywords}: drone traffic flow operations, differential game theory, decentralized control, multi-drone conflict resolution
\newpage

\section{Introduction}
Contemporary drone technology, including advanced monitoring and communication systems, will provide a plethora of applications, for instance in delivery and distribution, disaster relief, surveillance, but also personal mobility. As the demand for drone transportation increases (e.g.,  \cite{doole2020estimation}), multi-drone conflict resolution becomes more and more important, as is the efficient use of (scarce) airspace. We note that the term ``drone'' as used in this paper encompasses unmanned VTOL vehicles of varying sizes and weights, including eVTOL vehicles proposed for urban air mobility (UAM) services. We recognize that airspace regulations differentiate between vehicles of varying sizes (e.g. delivery drones are expected to operate at much lower altitudes than multi-person UAM vehicles). The approach in this paper is intended as a generic approach that does not make such distinctions, but might be applicable to any and all subsets of such vehicles.

\subsection{Multi-drone conflict resolution schemes}
In the paper, we focus on the topic of multi-drone conflict resolution schemes. Multi-drone conflicts are generally seen as problematic, and many approaches aim to avoid situations in which these conflicts occur. Our view is that as the demand for airspace is increasing, multi-drone conflicts cannot be prevented and approaches allowing us to efficiently deal with them are needed. 

An excellent overview of current conflict resolution (CR) schemes is given in \cite{ribeiro2020review}. Here, a distinction is made between centralized and decentralized schemes regarding both surveillance and trajectory propagation. In the review, explicit attention is paid to the assumptions of predictability of the paths of the opponent drones. 

With the expected large number of drones, we argue that centralized planning of all drone trajectories is not feasible nor practical, and we need to refer to decentralized conflict resolution, also called ``free flying'' \cite{Hoe:1998} in unstructured airspace. This forms the basis for the approach presented in this paper. 

We argue that the scientific foundation of many of the relevant concepts lies in traffic flow theory, as has been developed for (first) car traffic and (later) pedestrian traffic. Later in the paper, we will show some of the concepts used from that argument. Here, we will introduce the scientific background of applying traffic flow theory to air traffic. 

The authors of \cite{sunil2018three} show how air traffic can be operated if there are no specific flight paths.  This can be further structured as shown in \cite{sunil2016influence}. A further insight is that especially in urban environments, some of the low-altitude airspace is not available due to buildings. In that case,  airspace capacity is therefore limited by ``intersections''. First studies on this, both with simulations and analytically have been analyzed (e.g., \cite{aarts2023capacity}.  Alternatively, dedicated space is available for various directions. That can be in layers \cite{doole2021constrained} or in tubes \cite{cummings_2022}. Also a fundamental diagram, or macroscopic fundamental diagram, for has been shown for air traffic by simulation \cite{Cum:2021}. A resulting control measure, a form of perimeter control, has subsequently been presented and tested \cite{haddad2021traffic}.
In \cite{Ribeiro2022} reinforcement learning is used to optimize drone interaction rules to improve overall drone traffic operations. This is a great example of how we can use decentralized schemes to achieve efficient and safe flow operations. 
 
\subsection{Applications of game theory}
This paper introduces control concepts for drones using differential game theory. The approach is based on earlier approaches developed by the authors for connected vehicles \cite{Liu2022}, vessels \cite{Hoogendoorn201345}, pedestrians \cite{Hoogendoorn2003153}, and bicycle modeling \cite{Hoogendoorn2021}. The motivation for applying these techniques is the emergent phenomena that these decentralized approaches result in that we expect to be beneficial for drone traffic operations. 

Differential game theory has been applied to the control of drones before, but to the best knowledge of the authors not for real-time cooperative control purposes. Here, we propose a receding horizon approach in which each drone determines the best trajectory for the coming period while reconsidering this trajectory at chosen time instants. 

In the remainder, we will present the mathematical formulation of the differential game, which has been the focus of the presented research. We also present the numerical solution scheme and show some of the characteristics of the resulting decentralized multi-drone conflict resolution approach in terms of drone traffic flow operation characteristics. We also provide first insights into the sensitivity of these characteristics to the parameters of the decentralized drone control scheme. 

\section{Mathematical problem formulation}
In this section, we provide the mathematical problem formulation that forms the basis for the numerical solution scheme presented in the next section. We start by describing the simplified model capturing the drone dynamics, followed by the objective function and necessary conditions for optimality of the ego drone control. 

\subsection{A mathematical model for an ego drone in relation to its opponents}
We consider drone $i$ the behavior of which will be described by a receding horizon controller. The position of the drone is defined by a 3-dimensional vector $\vec{r}_i(t)$. The velocity of the drone and the acceleration of the drone are defined by $\vec{v}_i = {d \over dt} \vec{r}_i$ and $\vec{a}_i = {d \over dt} \vec{v}_i$ respectively. 

The dynamics of the ego drone are written down in the following state-space formulation:
\begin{equation}
    {d \over dt} \vec{x} = \vec{f} (t, \vec{x}, \vec{u})
\end{equation}
where the state $\vec{x}$ is defined by the positions and velocities of the ego drone $i$ and the other (opponents) drones $j \ne i$ with whom the ego drone interacts, subject to the given initial state $\vec{x}_i(t_k) = \vec{x}_k$. The vector $\vec{u}$ denotes the control signal - in this case, the acceleration vector $\vec{u}_i$. 

For this paper, we consider a very simple model to relate positions, velocities, and control signals. For the ego drone $i$ we have:
\begin{equation}
    {d \over dt} \vec{r}_i = \vec{v}_i
\end{equation}
\begin{equation}
    {d \over dt} \vec{v}_i = \vec{u}_i
\end{equation}
The same applies to the opponent drones $j$. 

We emphasize that more advanced models for the dynamics of drones can be used, and the methods presented in the ensuing can be applied without loss of generality. Here, we have decided to keep the dynamics as simple as possible to keep the mathematics as simple as possible.

\subsection{Drone cost specification}
The game-theoretical concept proposed here is based on the idea that the ego drone minimises some cost function. We will use the following cost functional:
\begin{equation}
    J_i = \int_{t_k}^{t_k+T} e^{-\eta s} L(s,\vec{x}(s),\vec{u}(s))ds + e^{-\eta (t_k + T)} \phi (t_k + T, \vec{x} (t_k+T))
\end{equation}
where $L$ denotes the running costs and $\phi$ denotes the terminal cost; note that while we have omitted subscripts, both costs are specific for the ego drone; $\eta$ denotes the discount factor showing that costs are discounted over time; $T$ denotes the control (and prediction) horizon reflecting how far the drone will plan ahead in time. For the sake of simplicity, we will assume $\phi=0$ in the ensuing.

In assuming that the ego drone will minimize the cost during the planning period $[t_k,t_k+T)$, we determine:
\begin{equation}
    \vec{u}_{[t_k,t_k+T)}^* = \arg \min J_i(t,\vec{x}(t_k), \vec{u}_{[t_k,t_k+T)})
\end{equation}
In other words, the ego drone determines the optimal acceleration vector functional, which will be applied from $t_k$ onward. A new optimal acceleration function will be computed at a new time $t_{k+1} < t_k + T$. 

For this paper, we will use simple cost specifications to derive the desired behavior of the ego drone. To this end, we split up the running costs into three components: 1) the cost of straying from the desired direction and speed; 2) the cost (risk) of getting too close to other drones, and 3) the cost of acceleration. For more advanced specifications of the running cost, other components can be added without loss of generality.  We propose:
\begin{equation}
    L^{stray} = {1 \over 2} \left( \vec{v}_i^0 - \vec{v}_i(t) \right)^2
\end{equation}
\begin{equation}
    L^{prox} = \sum_{j \ne i} e^{-d_{ij} / d_0}
\end{equation}
with $d_{ij} = || \vec{x}_j - \vec{x}_i||$, and
\begin{equation}
    L^{accel} = {1 \over 2} \vec{u}_p \cdot \vec{u}_p
\end{equation}

For combining the three factors, we propose a straightforward weighted linear combination of the three cost components:
\begin{equation}
\label{eq:1}
    L = \alpha L^{stray} + \beta_0 L^{prox} + L^{accel}
\end{equation}
where $\alpha$ and $\beta_0$ are weights. 

Next to opponent drones, in many situations obstacles also need to be considered. Here, we assume that the location of obstacles (walls, ground, etc.) are known. The obstacles are treated in the same way as the opponent drones: costs $L^{obs}$ are determined for each obstacle $k$
\begin{equation}
    L^{obs} = \sum_{k} e^{-d_{ik} / d_1}
\end{equation}
Here, $d_{ik}$ is the minimum distance from the ego drone $i$ to the obstacle $k$. The cost of being close to obstacles is added to the total cost specified in Eq. (\ref{eq:1}) by adding $\beta_1 L^{obs}$.

\subsection{Mathematical model of the opponent drones}
In the preceding subsection, we have formulated the optimization problem that will yield the  behaviour of the ego drone $i$, conditional on the behaviour of the opponent drones $j$. 
Note that we assume that $i$ knows the expected positions and velocities of all opponents for the prediction horizon and optimises its flying behaviour accordingly. At the same time, the opponents will behave in the same way, that is, optimise their behaviour under the assumption that they know the behaviour of the other drones.
The resulting problem is a {\em differential game}, where we are looking for the Nash equilibrium. In the remainder, we will show how we can determine this equilibrium state.

\section{Solving the differential game}
Now that the mathematical problem has been specified, let us briefly look at the solution approach. We describe the approach in two steps: the first step derives the necessary conditions for optimality of the solution using Pontryagin's minimum principle. The second step shows how these are used to determine the optimal solution using an iterative approach.

\subsection{Necessary conditions for optimality}
The differential game can be solved with the aid of
Pontryagin's minimum principle. To do this, we first need to determine the co-state dynamics. First, we define the Hamiltonian $H$ as follows:
\begin{equation}
    H = e^{-\eta t} L + \vec{\lambda} \cdot \vec{f}
\end{equation}
where $\vec{\lambda}$ denotes the shadow costs (or co-state) of the state $\vec{x}$. This co-state describes the relative
change in the optimal cost due to a small change in the state.

The optimality conditions provide the necessary conditions for the optimal acceleration:
\begin{equation}
    H(t,\vec{x}, \vec{u}^*, \vec{\lambda}) \le 
    H(t,\vec{x}, \vec{u}, \vec{\lambda})
\end{equation}
for all $\vec{u}$. For the model specification used in this paper, we can easily show that the optimality condition provides the following expression:
\begin{equation}
\vec{u}^* = -\vec{\lambda}_v    
\end{equation}
This shows that the optimal acceleration is equal to minus the marginal cost of the velocity $\vec{v}$. In other words, the ego drone applies the acceleration which yields the steepest reduction in the cost of the velocity. 

Pontryagin's Principle states that the co-states $\vec{\lambda}$ satisfy the following necessary conditions:
\begin{equation}
    -{d \over dt} \vec{\lambda} = {\partial \over \partial \vec x} H 
\end{equation}
For the co-states $\vec{\lambda}_v$ we have:
\begin{equation}
    -{d \over dt} \vec{\lambda}_v = 
    \alpha (\vec{v}^0 - \vec{v}_i)
    + \vec{\lambda}_r
\end{equation}
and for the co-states $\vec{\lambda}_r$ we have:
\begin{equation}
    -{d \over dt} \vec{\lambda}_r = 
    -{\beta \over d_0} \sum_{j \ne i} e^{-d_{ij} / d_0}
    \vec{n}_{ij} 
\end{equation}
Instead of initial conditions, which are given for the states, co-states satisfy terminal conditions that can be determined using the so-called transversality conditions:
\begin{equation}
    \vec{\lambda} (t_k+T) = {\partial \over \partial x} \phi(t_k + T, \vec{x}(t_k+T))
\end{equation}

This in principle provides a complete system of unknowns and equations: we have state dynamics with initial conditions, co-state dynamics with terminal conditions, and optimality conditions providing an expression for the optimal acceleration of the ego drone as a function of the co-state. This leaves us with a system of ordinary differential equations with mixed initial and terminal boundary conditions. 

\subsection{Iterative numerical solution}
In this section, we briefly discuss the iterative numerical solution approach. The algorithm is shown for one prediction period only; the receding horizon generalization is straightforward and left to the reader. Moreover, for the sake of simplicity, we have omitted obstacles, and terminal costs.  

\begin{enumerate}
    \item Initialization of control variables (prediction horizon $T$, time step $h$; 
    \item Initialization of parameters (weights, desired speed; relaxation parameter $a$, cut-off error $eps$
    \item For each drone, initialization of initial position $\vec r(0)$ and velocities $\vec v (0)$ and target position $\vec r_\infty$
    \item Initialize co-states for the positions $\vec \Lambda_r (t) = \vec 0$ and velocities $\vec \Lambda_v (t) = \vec 0$ for all $t=0:\Delta t: T$
    \item While $error > eps$ do
    \begin{enumerate}
        \item Set $\vec \lambda_r (t) = \vec \Lambda_r(t)$ and $\vec \lambda_v (t) = \vec \Lambda_v(t)$ 
        \item For $t$ = $0: \Delta t : T- \Delta t$
        \begin{enumerate}
            \item For $i = 1:n$
            \begin{enumerate}
                \item $\vec u(t|i) = - \vec \lambda_v (t|i)$
                \item $\vec v(t+\Delta t|i) = \vec v(t|i) + \Delta t \cdot \vec u (t|i)$
                \item $\vec x(t+\Delta t|i) = \vec x(t|i) + \Delta t \cdot \vec v (t|i)$
            \end{enumerate}
        \end{enumerate}
        \item For $t$ = $T: -\Delta t : \Delta t$
        \begin{enumerate}
            \item For $i = 1:n$
            \begin{enumerate}
                \item Compute desired velocity $\vec v_i^0(t)$
                \item $\vec{\lambda}_r (t-\Delta t|i) =  \vec{\lambda}_r (t|i)  + \Delta t \cdot {\beta \over d_0} \sum_{j \ne i} e^{-d_{ij} / d_0}  \vec{n}_{ij}$ 
    \item $\vec{\lambda}_v (t-\Delta t|i) = \vec \lambda_v (t|i) + \Delta t \cdot \left( \alpha (\vec{v}_i^0 - \vec{v} (t|i))
    + \vec{\lambda}_r (t|i) \right)$
      \end{enumerate}
            \end{enumerate}
            \item Relaxation $\vec \Lambda_r(t) = (1-a) \cdot  \vec \Lambda_r(t) + a \cdot \lambda_r(t) $ and $\vec \Lambda_v(t) = (1-a) \cdot  \vec \Lambda_v(t) + a \cdot \lambda_v(t) $
            \item $error = || \vec \Lambda - \vec \lambda ||$
        \end{enumerate}
    \end{enumerate}

\begin{figure}[ht]
\centering
\includegraphics[height=8cm]{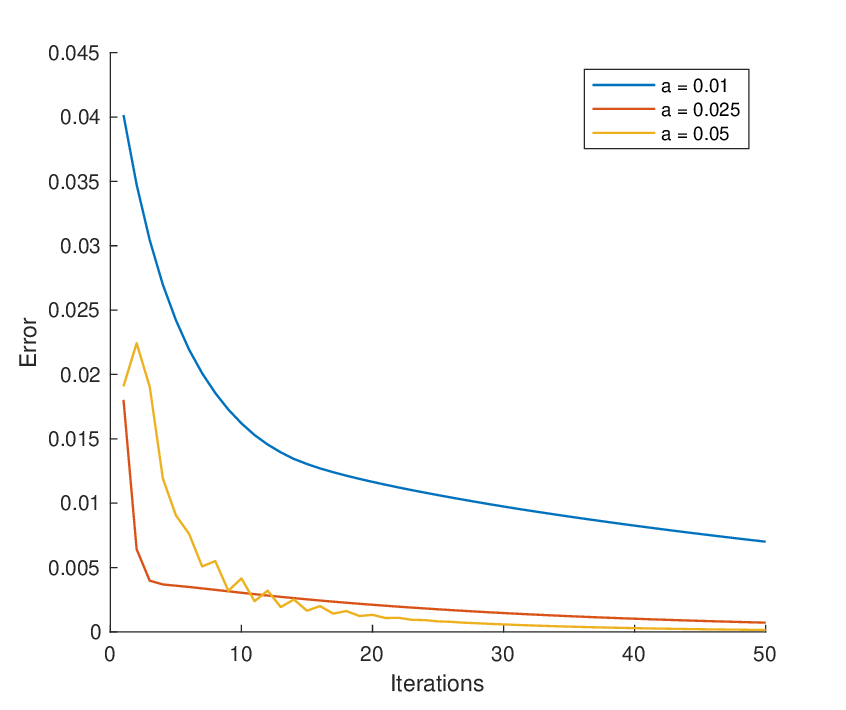}
\caption{Convergence behavior of iterative scheme for 50 iterations and different values of $a$}
\label{fig:convergence}       
\end{figure}

It is beyond the scope of the paper to analyze the performance of the numerical solution in detail. For illustration purposes, Fig. \ref{fig:convergence} shows the convergence properties of the scheme for a one-on-one drone interaction scenario, with a time horizon of 10 s (discussion in the following section for more details) for different values of the relaxation parameter $a$. The picture clearly shows how $a$ affects the rate of convergence. We also note that for this case, larger values of $a$ lead to non-convergence of the scheme (not shown in the figure). Convergence issues already occur for $a=0.05$, where oscillations are observable due to over-correction of the stable solution $\vec{\Lambda}$. While further investigation into the convergence properties is an important direction for future research, it is out of the scope of this paper. 

\section{Experiments}
In this section, we present the results of applying the multi-drone conflict resolution approach presented in this paper to different scenarios. We emphasize that the objective is to gain first insights into the properties of the scheme first and foremost, and not to present real-life scenarios. This implies that results would need to be scaled (in terms of time and spatial scale) to realistic values upon true-life application. Moreover, the number of scenarios presented is limited as they serve to illustrate the emerging properties of the multi-drone conflict resolution scheme. 

\subsection{One-on-one drone interaction}
\label{sec:one-on-one-exp}
\begin{figure}[ht]
\centering
\includegraphics[height=9cm]{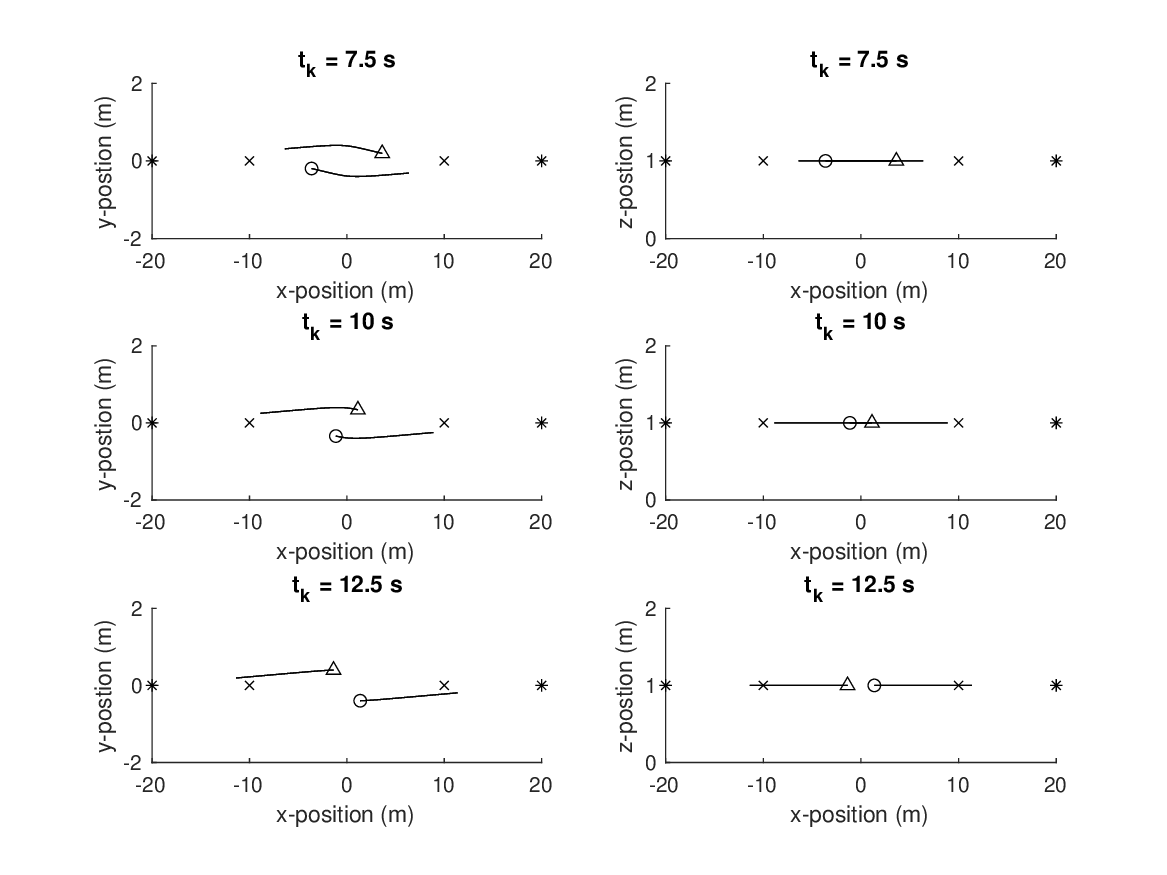}
\caption{Example of model behavior for head-on interaction between two drones for $v^0 = 1$, $T = 10$, $\alpha = 1$, and $\beta_0 = 10$.}
\label{fig:1}       
\end{figure}

Fig. \ref{fig:1} shows results from solving the game theoretical problem for a one-on-one drone conflict. The situation shows how two drones try to reach opposite destinations (the asterix '*'). The figure shows the planned trajectory from the current position at $t_k$. The planned trajectory is the trajectory that minimizes the cost of the ego drone. The figure shows the situation as a projection on the $x-y$ plane and the $x-z$ plane.  
Note that in this example, the interaction takes place in plane $z=1$. This is not necessarily the case, but a consequence of the slightly perturbed initial conditions in the $x$ and $y$ position of the drones. In fact, there are infinite optimal solutions for a head-on interaction as considered in this scenario: the algorithm will compute only one. 

To test the influence of the different parameters numerically, have considered the head-on drone interaction scenario for different parameter values. For each of the considered scenarios, the impacts of the speeds of the drones turned out to be negligible. We therefore only show the impact of the minimum distance between the passing drones in relation to the parameter values. Table \ref{tab:1} shows an overview of the impacts of some of the key parameters that can be used in designing the drone scheme. 

First of all, the prediction (or look-ahead) time $T$ determines the minimum distance between the two drones (in a non-linear way). The table shows that the larger the look-ahead time, the larger the minimum distance between the two drones is. This is in line with our expectations, as looking further ahead allows for earlier changes in the direction and speed of the drone.

Second, the interaction scaling parameter $R$ has been considered. Its impact is also as expected: the minimum distance scales approximately linearly with $R$. The parameter hence provides a good way to design the minimum distance allowed between two interacting drones.  

Finally, the table shows the impact of the weight $\beta_0$ of the proximity costs. Again, the parameter influences the minimum distance as expected: a larger value yields more emphasis on maintaining sufficient distance at the cost of changing course and speed. 

\begin{table}[ht]
    \centering
    \begin{tabular}{c c c}
    \hline
         Parameter & Value  & Minimum distance \\
         \hline
         $T$ & 2.5 & 0.21 \\
         & 5 & 0.34 \\
         & 10 & 0.41 \\
         \hline
         $R$ & 0.05 & 0.18 \\
         & 0.1 & 0.34 \\
         & 0.2 & 0.68 \\
         \hline
         $\beta_0$ & 1 & 0.16 \\
         & 10 & 0.34 \\
         & 100 & 0.55 \\
         \hline
    \end{tabular}
    \caption{Minimum distance sensitivity for head-on scenario with $v^0 = 1$, $\alpha = 1$, $\beta_0 = 10$, $T=5$, and $R=0.1$; and $\Delta t = 1 s$}
    \label{tab:1}
\end{table}

\subsection{Self-organization in multi-drone conflict resolution }
In this subsection, we consider multi-drone conflict resolution scenarios, focusing on the self-organization characteristics of the scheme. To this end, we look at a head-on interaction scenario and a crossing scenario. We use the same parameter values as in the one-on-one interaction case unless stated differently.  

\subsubsection{Head-on interactions and lane formation}
The first multi-drone scenario considered is a scenario where two groups of drones are meeting head-on. To this end, we generated two groups of drones in two rectangular boxes of 5 by 2 by 1 meter around $x=12.5$ and $x=-12.5$, $y=0$ and $z=1$ respectively. In each box, $n$ drones are generated. For the right box, drones move towards the left (desired direction $\vec e^0 = (-1,0,0)$), while for the left box, drones move towards the right (desired direction $\vec e^0 = (1,0,0)$). The locations of the $n$ drones are drawn from a 3D uniform distribution, assuming equal mean distances in each direction. The initial speeds of the drones are equal to zero. 

\begin{figure}[ht]
\centering
\includegraphics[height=9cm]{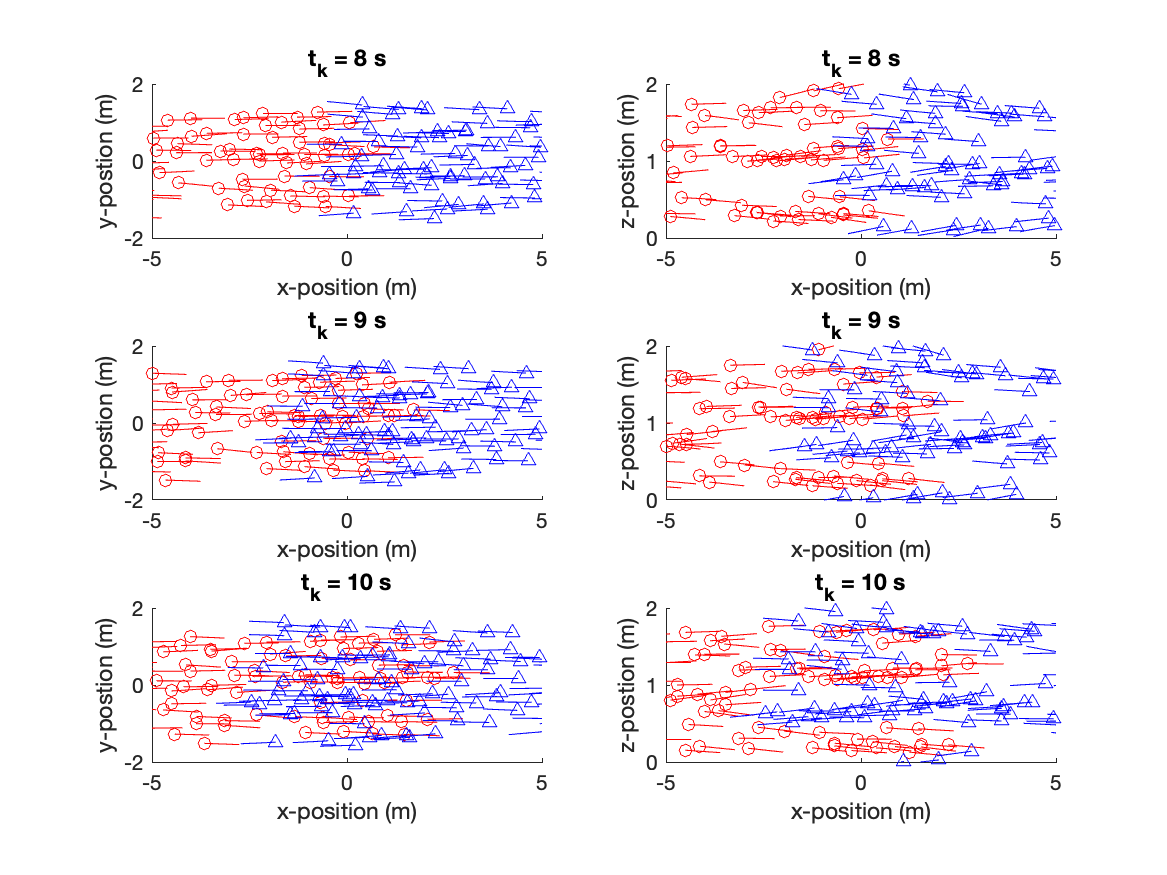}
\caption{Example showing self-organization in bi-directional drone flow operations for head-on scenario (100 drones) with $v^0 = 1$, $\alpha = 1$, $\beta_0 = 10$, $T=5$, and $R=0.1$; and $\Delta t = 1$}
\label{fig:bi-flow-example1}       
\end{figure}

To illustrate the performance of our decentralized approach, we will perform a range of analyses and show different visualizations. 
To start, Fig. \ref{fig:bi-flow-example1} illustrates the results of the head-on interaction experiment, by showing three consecutive snapshots of the drone positions and their intended paths (for 1 s ahead). The figure shows the projection on the x-y plane (left column of pictures) and on the x-z plane (right column of pictures). 
Although more difficult to discern than in the case of a 2D pedestrian flow, we can clearly see that clusters are formed by drones that are moving in the same direction. This cluster-formation resembles the formation of bi-directional lanes in a pedestrian flow \cite{Helbing20001240}, but is more complex as the drones have more freedom to move. As a result, patterns are not always easily identifiable. 

Next, let us consider the result statistics shown in Tab. \ref{tab:2}, showing the impact of the parameters of the algorithm on the minimum distance and the average speed \textit{in the direction of the destination}. Regarding the latter, the table shows that the speed is not strongly affected by the parameters, but the minimum distance observed is. This shows that the scheme results in very efficient multi-drone conflict resolution. 

For the prediction horizon $T$, we see that the minimum distance increases when the prediction horizon becomes larger. This is expected as looking ahead further improves the possibilities for more efficient interactions with other drones. The improvement reduces substantially when going from $T=5$ to $T=10$, at the expense of increased computational complexity. For this case, we could argue that $T=5$ is a good trade-off between efficiency and complexity. 

The minimum distance scales linearly with the distance scaling $R$. Similar to the one-on-one interaction, this shows that $R$ is an important design parameter that directly influences interaction characteristics. To an extent, the same applies to the parameter $\beta_0$: increasing the weight of the proximity costs results in a higher importance of keeping sufficient distance between the interacting drones, at the expense of a slight reduction in efficiency. 

\begin{table}[ht]
    \centering
    \begin{tabular}{c c c c}
    \hline
         Parameter & Value  & Minimum distance & Average speed\\
         \hline
         $T$ & 2.5 & 0.33 & 0.99\\
         & 5 & 0.44 & 1.00\\
         & 10 & 0.46 & 1.00 \\
         \hline
         $R$ & 0.05 & 0.22 & 1.00\\
         & 0.1 & 0.42 & 1.00\\
         & 0.2 & 0.85 & 0.99 \\
         \hline        
         $\beta_0$ & 1 & 0.27 & 1.00\\
         & 10 & 0.47 & 1.00\\
         & 100 & 0.59 & 0.99 \\
         \hline
    \end{tabular}
    \caption{Minimum distance and average speed sensitivity for head-on scenario (100 drones) with $v^0 = 1$, $\alpha = 1$, $\beta_0 = 10$, $T=5$, and $R=0.1$; and $\Delta t = 1$}
    \label{tab:2}
\end{table}

Overall, we can conclude that the interactions are highly efficient, even in the case of high numbers of interacting drones. This has two reasons. First, as the interactions occur in free space, the swarm expands as more drones are present. Later in this section, we will consider restricted airspace to showcase the impact on efficiency by introducing bottleneck scenarios. 
Second, as we have already seen by looking at a single simulation outcome, self-organized patterns are formed that enable efficient interaction of the drone groups. 
 
\begin{figure}[ht]
\centering
\includegraphics[height=9cm]{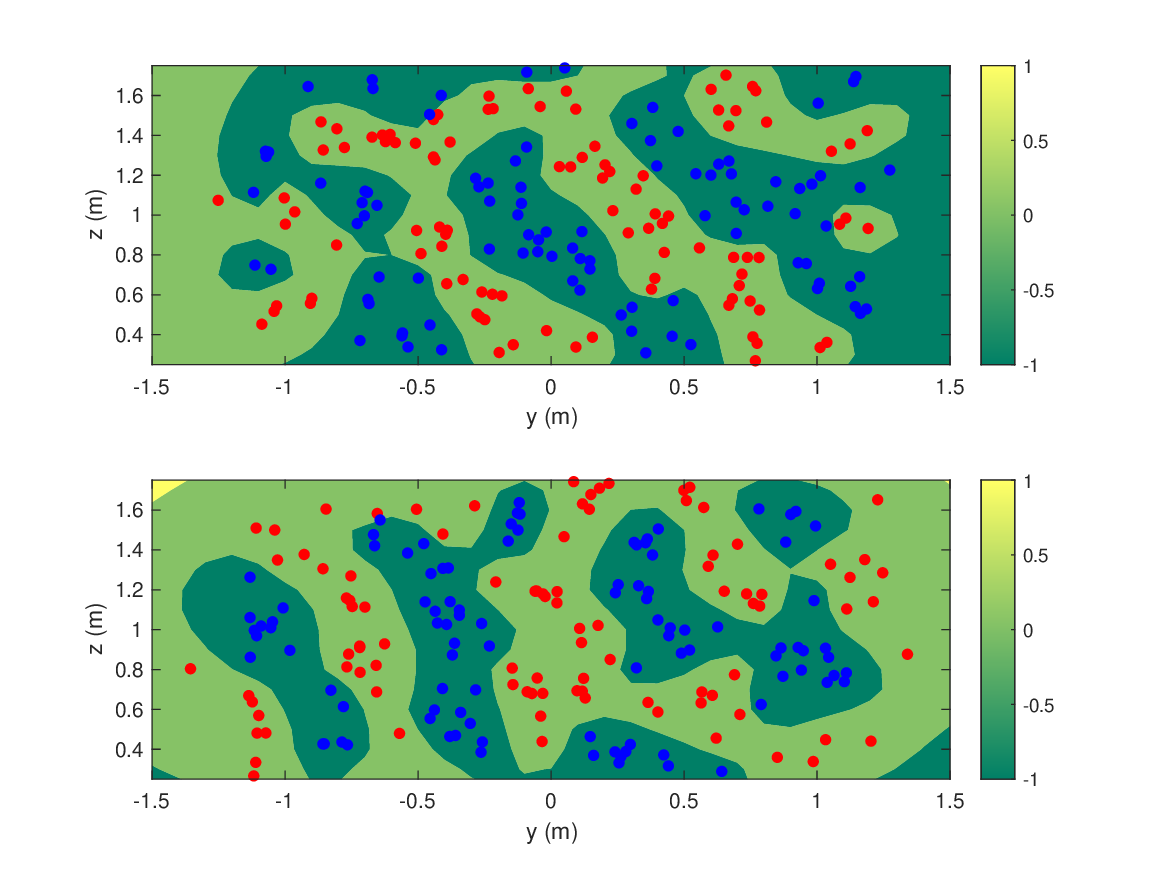}
\caption{Two examples of the spatial distribution of drone groups for head-on interaction scenario near $x=0$. The two pictures show two random simulation outcomes from the same scenarios, resulting from different initial spatial distributions of the drones.}
\label{fig:cluster}       
\end{figure}

To visualize the self-organized patterns, we consider the spatial distribution of the drones from the two groups near the point where the head-on interaction occurs (i.e., at $x=0$). To this end, we collected the points at which the drones of both groups passed the $x=0$ cross-section. 
The picture shows these points clustered based on direction. For the clustering, the 'natural' interpolation function for irregularly gridded data of Matlab 2023b was used (for visualization purposes only). 
Fig. \ref{fig:cluster} shows the patterns that are formed for a 200-drone interaction experiment. The figure clearly shows the clusters that are formed for the two example situations, where we emphasize that the points shown reflect different passing times. Clearly, the scheme results in a structure where drones moving in the same direction form groups that are spatially and temporarily close, as to reduce the number of close-range interactions with drones moving in the opposite direction. 

\subsubsection{Crossing flows and self-organized patterns}
Here, we briefly discuss the results of a crossing drone flow experiment. Again, two groups of drones are considered. Instead of meeting heads-on, the drone flows cross at a 90-degree angle. Group 2 moves along the x-axis (from right to left), while group 1 moves along the y-axis (from top to bottom). In line with the formation of diagonal striped patterns in pedestrian flow (see \cite{Helbing20001240}), we expect specific patterns to be formed for this scenario that resemble structures formed in pedestrian flow operation; see \cite{Helbing20001240}.

\begin{figure}[ht]
\centering
\includegraphics[height=11cm]{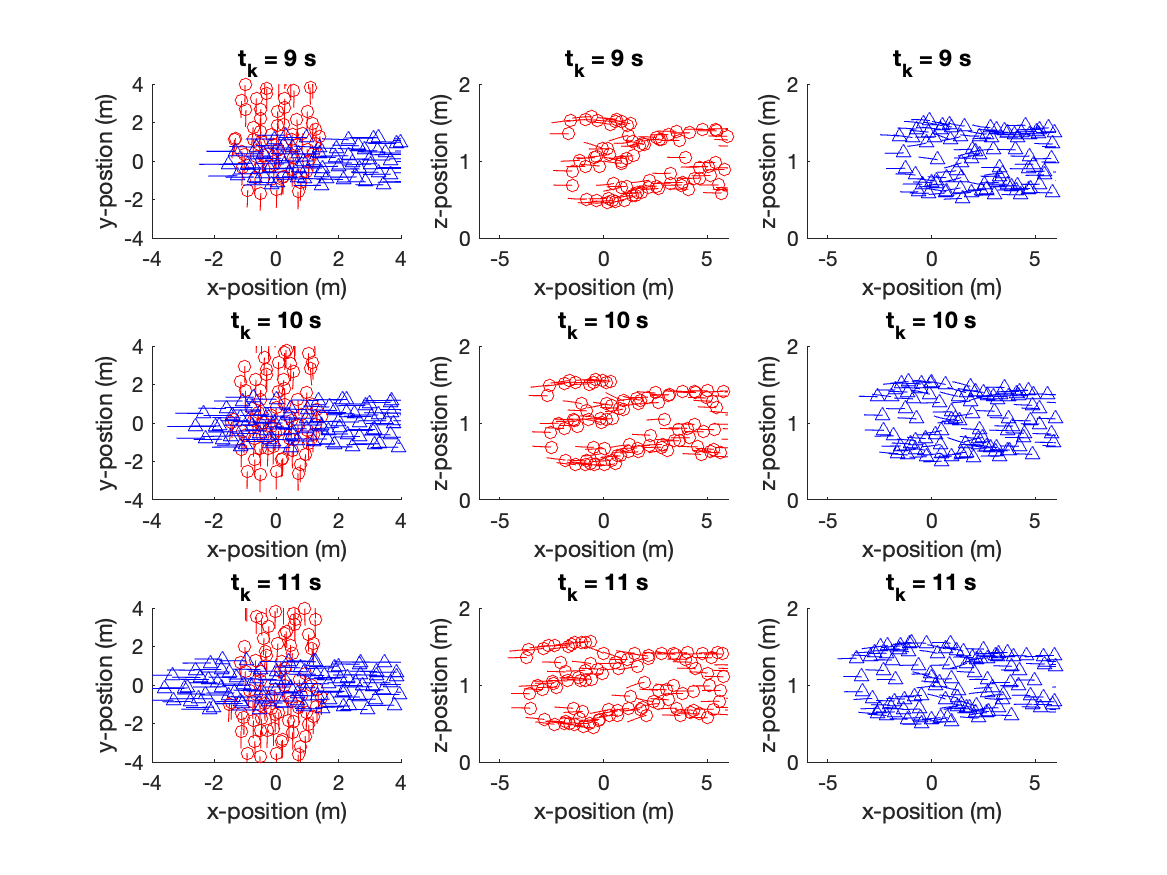}
\caption{Illustration of drone behavior for crossing experiment 1. The left column shows the projections of the drone positions on the $x-y$ plane; the middle and left columns show the projections of groups 1 and 2 on the $x-z$ and $y-z$ plane respectively. We plotted the current position, and the 1-second ahead planned trajectory.}
\label{fig:crossing1}       
\end{figure}

\begin{figure}
\centering
\includegraphics[height=11cm]{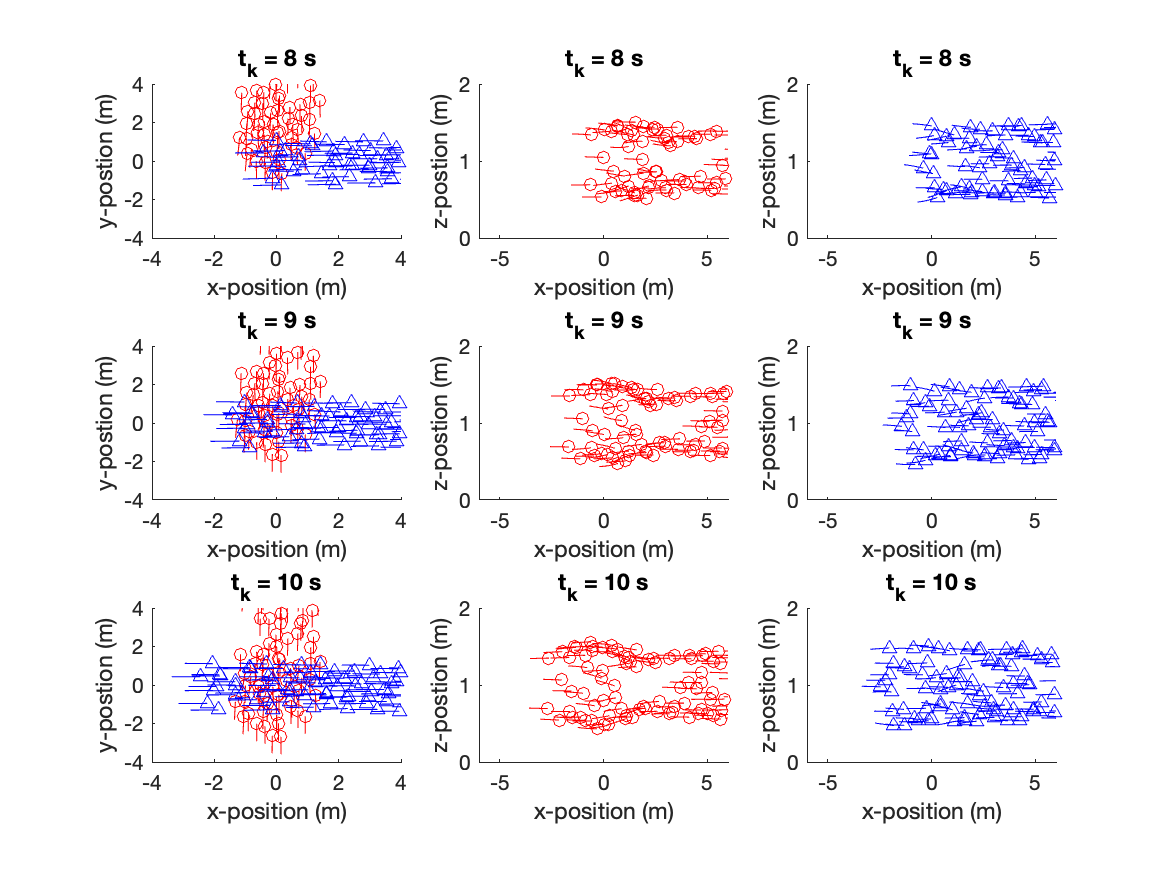}
\caption{Illustration of drone behavior for crossing experiment 2.}
\label{fig:crossing2}       
\end{figure}

Figures \ref{fig:crossing1} and \ref{fig:crossing1} show the results of two randomly chosen experiments, where random means that, in line with the head-on interaction case, initial positions of the 200 drones were randomly generated in two bounding boxes. Both pictures show that forms of self-organization are present. While the diagonal patterns, which we know from pedestrian flow operations, are not directly transferred to the 3D drone flow, we do see similar emergent patterns. For instance, in Fig. \ref{fig:crossing1}, we see that group 1 forms two diagonal shapes in the z-direction. The patterns formed in Fig. \ref{fig:crossing2} are different, yet clusters are clearly identifiable. 

To gain insight into the sensitivity of the efficiency and the risk levels of the flow operations of the parameter values, we repeated the sensitivity analysis of the head-on scenario to the crossing flow scenario. Tab. \ref{tab:3} shows the result of this sensitivity analysis for the crossing flow scenario. 

Let us first look at the impact of the prediction horizon $T$. Interestingly, we see that while increasing $T$ from 2.5 to 5 yields an increase in the minimum distances, increasing it further causes a reduction (i.e., an increase in the risk). This counterintuitive outcome shows that what is optimal from the perspective of an individual drone, may not per se yield a system improvement for specific cases. While further investigations are needed, we do remark that similar results are reported in \cite{Hoogendoorn2021}. 

Looking at the impact of the interaction distance $R$, we see similar results as in the head-on scenario. First of all, it is remarkable to see that average speeds are insensitive to the value of $R$, meaning that the increased spacing requirements have a very limited impact on the average operation speed of the drones. It again needs to be stressed that in this scenario, there are no spatial constraints and airspace capacity is - in a sense - unlimited. From the table, we can clearly see how the minimal distance between the drones scales with increasing $R$: as with the head-on scenario, $R$ can be effectively used to ensure that drone operations meet safety requirements expressed in the distances the drones maintain from each other.  
Again, the same applies to the parameter $\beta_0$: as for the head-on scenario, increasing the weight of the proximity costs results in a higher importance of keeping sufficient distance between the interacting drones. 

\begin{table}[ht]
    \centering
    \begin{tabular}{c c c c}
    \hline
         Parameter & Value  & Minimum distance & Average speed\\
         \hline
         $T$ & 2.5 & 0.31 & 0.99\\
         & 5 & 0.43 & 0.99\\
         & 10 & 0.37 & 1.00 \\
         \hline
         $R$ & 0.05 & 0.27 & 0.99\\
         & 0.1 & 0.43 & 0.99\\
         & 0.2 & 0.80 & 0.99 \\
         \hline        
         $\beta_0$ & 1 & 0.22 & 1.00\\
         & 10 & 0.45 & 0.99\\
         & 100 & 0.47 & 0.99 \\
         \hline
    \end{tabular}
    \caption{Minimum distance and average speed sensitivity for crossing-flow scenario (200 drones) with default settings $v^0 = 1$, $\alpha = 1$, $\beta_0 = 10$, $T=5$, and $R=0.1$; and $\Delta t = 1$}
    \label{tab:3}
\end{table}

\subsection{Bottlenecks and multi-drone conflict resolution}
In this subsection, we consider multi-drone conflict resolution scenarios for bottleneck scenarios. To this end, we reconsider the head-on interaction scenario discussed in the previous subsection. In this case, however, we restrict the amount of airspace available by positioning two large static obstacles with a radius of $2m$ located at $(x,y)$ $(0,-2.5)$ and $(0,2.5)$ respectively. While the drones are able to fly around the obstacles (and may, in particular for longer prediction horizons), the shortest path is to navigate between the 1-meter passage between the cylinders. 

\begin{figure}
\centering
\includegraphics[height=11cm]{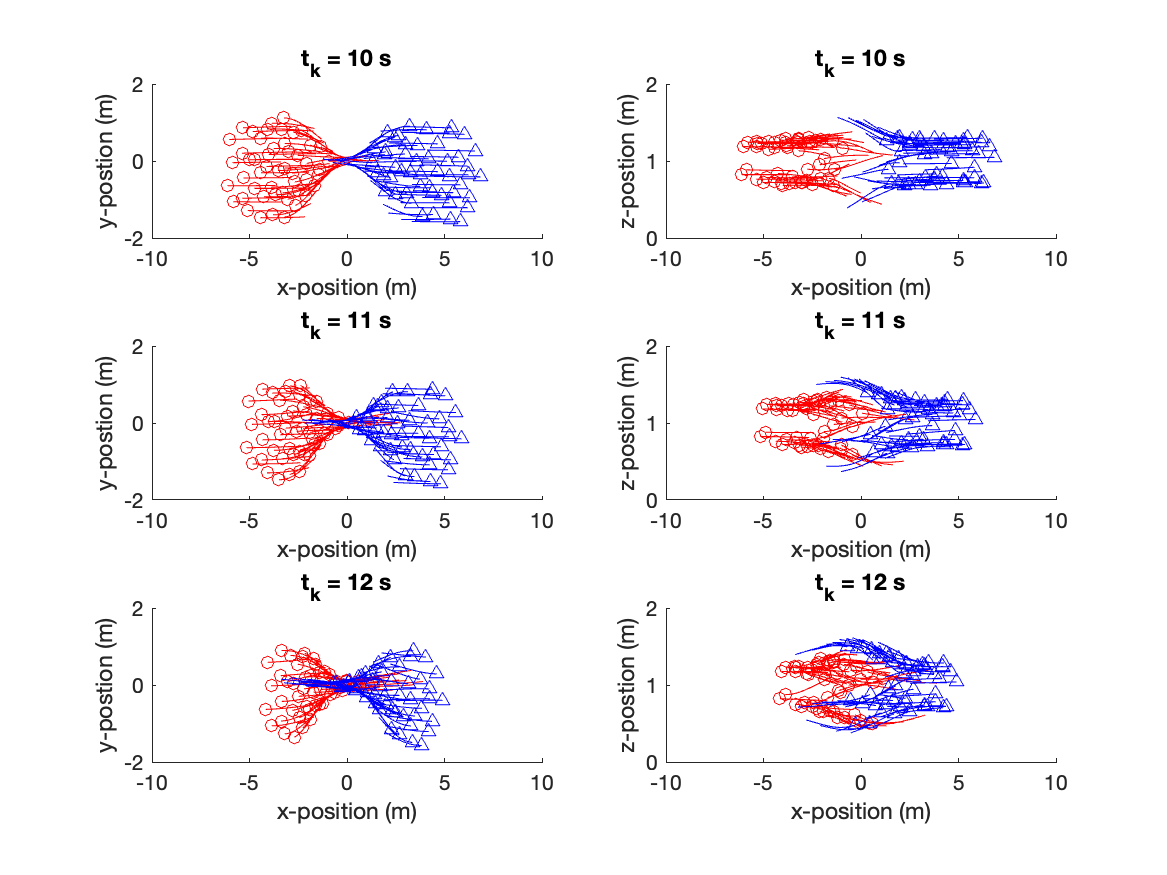}
\caption{Illustration of cluster formation process inside the bottleneck; left pictures show top view; right pictures show side view.}
\label{fig:clusterformation}       
\end{figure}

Also in this scenario, we see self-organization occurring, leading to efficient flow operations. Fig. \ref{fig:clusterformation} showcases this by looking at the operations in three consecutive time instants. The pictures show the current position of the drones (circle or triangle) and the planned optimal trajectory determined using the multi-drone conflict resolution scheme presented in this paper. Note that we have used $T=2.5s$ to make pictures somewhat clearer. The figure shows that before the drones arrive at the bottleneck, they have formed two clusters for each direction. When the drones arrive at the bottleneck, the flocks funnel to form a narrow stream, which is clearly observable from the top view (left column of pictures). At the same time, they start to form dynamic lanes when navigating through the bottleneck. The separation between the two directions occurs in the $z$-direction mostly, although in other simulations (and with other parameters) other patterns occur as well. 

\begin{figure}
\centering
\includegraphics[height=9cm]{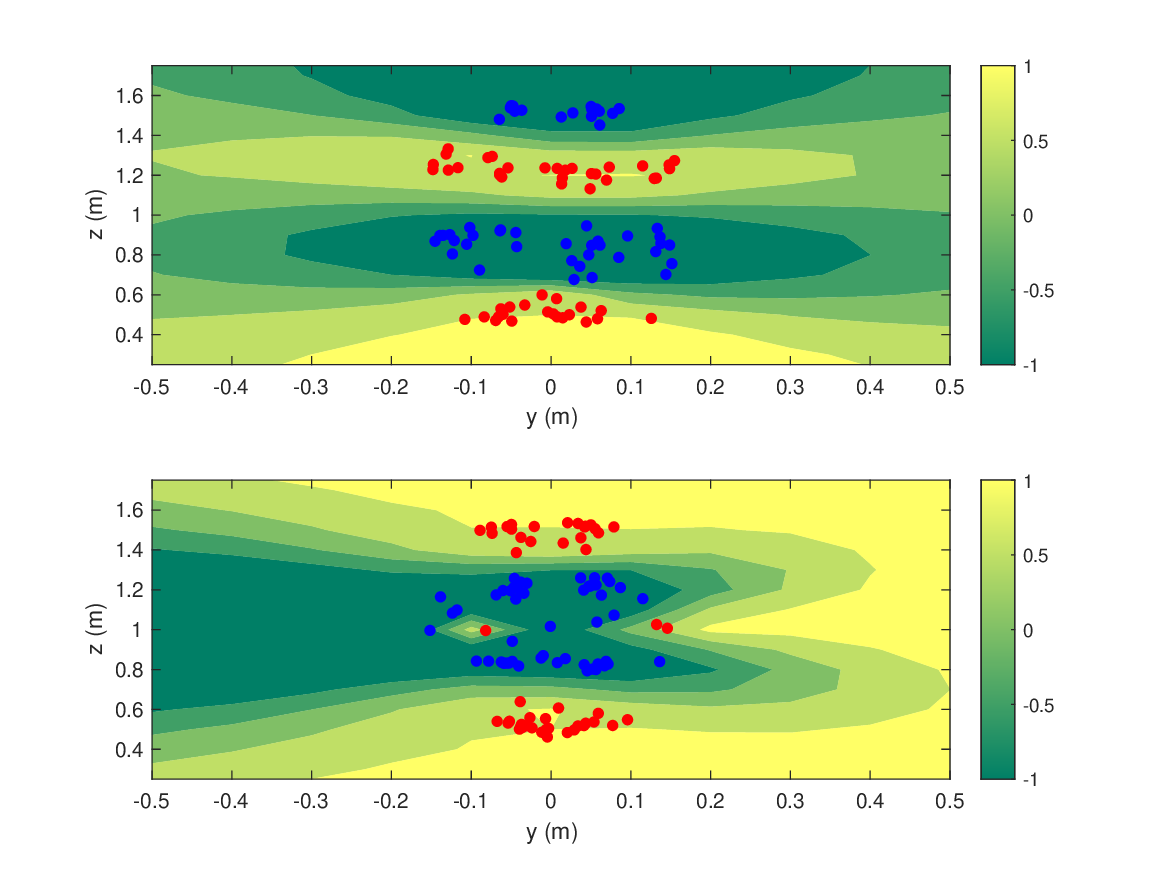}
\caption{Two examples of the spatial distribution of drone groups for the bottleneck scenario near $x=0$. The two pictures show two random simulation outcomes from the same scenarios, resulting from different initial spatial distributions of the drones.}
\label{fig:clusterbottle}       
\end{figure}

Fig. \ref{fig:clusterbottle} shows the self-organization in a different way, by providing the cross-sectional projection of the passing drones at $x=0$ for two random simulation experiments. It again shows how the bi-directional flows are separated in the z-direction. It also shows how in different situations, a different number of lanes are formed (top: 4 lanes, bottom: 3 lanes). 

To get some insight into the impact of the parameter values, Tab. \ref{tab:4} shows the sensitivity of the minimum distance (as a measure for risk) and average speed (as a measure of efficiency) to different parameter settings (i.e., $R$ and $\beta_0$). 
From the table, we can conclude that the parameters have the expected results. First of all, we see that changing the distance scaling parameter $R$ impacts the minimum distance in an approximate linear fashion. While the minimum distance that drones maintain increases with increasing $R$, the average speed starts to reduce. This is caused by the fact that due to increased spatial requirements, bottleneck capacity is insufficient and drones need to reduce their speed before being able to pass through the bottleneck (congestion). 
The impact of the weighting parameter $\beta_0$ is also as expected given the outcomes of the previous experiments. While increasing the weight increases the minimum distances between the drones, the average speed reduces signifying forms of congestion occuring due to the bottleneck.   

\begin{table}[ht]
    \centering
    \begin{tabular}{c c c c}
    \hline
         Parameter & Value  & Minimum distance & Average speed\\
         \hline
         $R$ & 0.05 & 0.22 & 0.98\\
         & 0.1 & 0.35 & 0.92\\
         & 0.2 & 0.67 & 0.73 \\
         \hline        
         $\beta_0$ & 1 & 0.17 & 0.99\\
         & 10 & 0.16 & 0.86\\
         & 100 & 0.24 & 0.62 \\
         \hline
    \end{tabular}
    \caption{Minimum distance and average speed sensitivity for head-on scenario (100 drones) with $v^0 = 1$, $\alpha = 1$, $\beta_0 = 10$, $T=5$, and $R=0.1$; and $\Delta t = 1$}
    \label{tab:4}
\end{table}

\section{Discussion, conclusions and future work}
This paper presents a novel approach to multi-drone conflict resolution based on differential game theory, which has been successfully used in modeling pedestrian, cyclist, and vessel dynamics, as well as controlling CAVs. The problem formulation and numerical solution approaches were adapted here for multi-drone conflict resolution.

Based on the results presented in this paper, we conclude that the proposed game-theoretical approach shows potential for real-time application in multi-drone conflict resolution and decentralized management of drone traffic. While showing promise, several aspects require further investigation before practical applications can be considered. Some of these required investigations are theoretical in nature, others are more technical in the sense that insights in (the limits of) communication and sensing technology. Below, we provide an overview of our findings as well as the required further studies.  

\subsubsection{Main findings}
First of all, the numerical properties of the solution scheme are sufficient to warrant further investigation: the scheme converges to solutions close to the optimal solution within a limited number of iterations, and computation times are acceptable for real-time application. However, further analyses are required, especially when considering realistic scenarios with accurate update times, prediction horizons, and interactions with other drones. Such analyses were out of the scope of this paper, which aimed to show the concept rather than the practical details of real applications.

Secondly, the emergent drone flow operations prove to be highly efficient while maintaining safety requirements. This efficiency is attributed to the self-organized structures that occur when different groups of drones interact. These patterns are similar to those observed in pedestrian and, to a lesser extent, bicycle flows, which motivated the application of differential game theory to drones. The experiments demonstrate that even in the presence of airspace restrictions, self-organized processes occur, resulting in efficient drone operations. However, in situations of limited capacity, average drone speeds decrease as queuing forms in front of the bottleneck. Furthermore, the limiting behavior of the system as density increases and conflicts correspondingly increase, losses in productivity are likely to be experienced providing an upper bound to the system's practical capacity, as illustrated in the fundamental diagrams observed in \cite{Cum:2021}.

\subsubsection{Further investigations}
From a scientific perspective, further investigation of the various phenomena observed in three-dimensional space is of key interest. One of the crucial issues is the limit of efficient and safe self-organization, including the role of heterogeneity in reducing this efficiency of self-organization. For pedestrian flow operations, we know that self-organization breaks down when densities become too high. Furthermore, heterogeneity, expressed in terms of the free speeds $\vec{v}_i^0$ of the drones, or other differences in drone performance characteristics, has a negative impact on self-organization leading to the so-called 'freezing by heating' effect that is observed in different self-driven systems. For instance, in \cite{Helbing20001240} it is shown how heterogeneity in pedestrian flows leads to failing self-organization of bi-directional lanes and consequently to blockages. The limits in self-organization will result in the need for more centralized interventions in terms of air traffic management, structuring the airspace, etc. These could pertain to the prioritization of specific types of drones, management measures such as speed homogenization, virtual intersection controllers, etc. 

Our enhanced insight into drone flow properties will also form the basis for aggregate (macroscopic) modeling approaches that may be suitable to capture the dynamics of swarms of drones driven by game-theoretical control principles. Understanding the conditions under which self-organization collapses can provide insights for designing drone traffic management control principles, including the redistribution of capacity to specific groups of drones if needed and possible airspace architectures in areas where restrictions are called for by land use and environmental considerations as in \cite{cummings_2023}. Hierarchical approaches have been successful in the past in the CAV domain \cite{Liu2022}.

The impact of sensing and communication errors needs to be studied. Different strategies proposed in \cite{Hoogendoorn2021} express the level of cooperation among traffic participants. We argue that risks arising from high probabilities of sensing and communication failure can be mitigated by considering the opponent drones as potentially hostile (the 'demon drone' scenario). This approach ensures that the ego drone adopts risk-averse strategies. At the same time, more complex drone dynamics models are to be investigated to understand the gains (in terms of e.g. reduction of model error) and losses (increased computation complexity) of including more involved drone dynamics. 

In addition to analyzing technical limitations, it will be crucial to consider compliance with existing or future drone traffic regulations (e.g., European regulations for airspace use, U-space), as well as priorities for different types of drones. These rules will often imply more structured airspace, where drones with specific functions or destinations will move in specific airspace layers. They furthermore may involve priorities for specific functions. 
Examples of how to include traffic rules in a differential game theoretical approach to conflict resolution and interaction modeling can be found in \cite{Hoogendoorn2021}.

\section{Author contributions}
The authors confirm their contribution to the paper as follows: study conception and design: Hoogendoorn;  simulation preparation and execution: Hoogendoorn; data analysis and interpretation of results: Hoogendoorn, Knoop, Mahmassani; draft manuscript preparation: Hoogendoorn, Knoop, Mahmassani, Hoogendoorn-Lanser. All authors reviewed the results and approved the final version of the manuscript.
 
\printbibliography

@ARTICLE{Liu2022,
author={Liu, M. and Zhao, J. and Hoogendoorn, S. and Wang, M.},
title={A single-layer approach for joint optimization of traffic signals and cooperative vehicle trajectories at isolated intersections},
journal={Transportation Research Part C: Emerging Technologies},
year={2022},
volume={134},
doi={10.1016/j.trc.2021.103459},
art_number={103459},
note={cited By 12},
document_type={Article},
source={Scopus},
}

@ARTICLE{Hoogendoorn2021,
author={Hoogendoorn, S. and Gavriilidou, A. and Daamen, W. and Duives, D.},
title={Game theoretical framework for bicycle operations: A multi-strategy framework},
journal={Transportation Research Part C: Emerging Technologies},
year={2021},
volume={128},
doi={10.1016/j.trc.2021.103175},
art_number={103175},
note={cited By 2},
document_type={Article},
source={Scopus},
}

@ARTICLE{Hoogendoorn201345,
author={Hoogendoorn, S. and Daamen, W. and Shu, Y. and Ligteringen, H.},
title={Modeling human behavior in vessel maneuver simulation by optimal control and game theory},
journal={Transportation Research Record},
year={2013},
number={2326},
pages={45-53},
doi={10.3141/2326-07},
note={cited By 11},
document_type={Article},
source={Scopus},
}

@ARTICLE{Hoogendoorn2003153,
author={Hoogendoorn, S. and Bovy, P.H.L.},
title={Simulation of pedestrian flows by optimal control and differential games},
journal={Optimal Control Applications and Methods},
year={2003},
volume={24},
number={3},
pages={153-172},
doi={10.1002/oca.727},
note={cited By 203},
document_type={Review},
source={Scopus},
}

@ARTICLE{Ribeiro2022,
author = {Ribeiro, Marta and Ellerbroek, Joost and Hoekstra, Jacco},
title = {Improving Algorithm Conflict Resolution Manoeuvres with Reinforcement Learning},
year = {2022},
journal = {Aerospace},
volume = {9},
number = {12},
doi = {10.3390/aerospace9120847},
publication_stage = {Final},
source = {Scopus},
note = {Cited by: 1; All Open Access, Gold Open Access}
}

@article{ribeiro2020review,
  title={Review of conflict resolution methods for manned and unmanned aviation},
  author={Ribeiro, Marta and Ellerbroek, Joost and Hoekstra, Jacco},
  journal={Aerospace},
  volume={7},
  number={6},
  pages={79},
  year={2020},
  publisher={MDPI}
}

@ARTICLE{Helbing20001240,
	author = {Helbing, Dirk and Farkas, Illés J. and Vicsek, Tamás},
	title = {Freezing by heating in a driven mesoscopic system},
	year = {2000},
	journal = {Physical Review Letters},
	volume = {84},
	number = {6},
	pages = {1240 – 1243},
	doi = {10.1103/PhysRevLett.84.1240},
	type = {Article},
	publication_stage = {Final},
	source = {Scopus},
	note = {Cited by: 374; All Open Access, Green Open Access}
}

@Article{haddad2021traffic,
  author    = {Haddad, Jack and Mirkin, Boris and Assor, Kfir},
  title     = {Traffic flow modeling and feedback control for future Low-Altitude Air city Transport: An MFD-based approach},
  journal   = {Transportation Research Part C: Emerging Technologies},
  year      = {2021},
  volume    = {133},
  pages     = {103380},
  publisher = {Elsevier},
}

@Article{sunil2018three,
  author    = {Sunil, Emmanuel and Ellerbroek, Joost and Hoekstra, Jacco M and Maas, Jerom},
  title     = {Three-dimensional conflict count models for unstructured and layered airspace designs},
  journal   = {Transportation Research Part C: Emerging Technologies},
  year      = {2018},
  volume    = {95},
  pages     = {295--319},
  publisher = {Elsevier},
}

@InProceedings{sunil2016influence,
  author    = {Sunil, Emmanuel and Hoekstra, Jacco and Ellerbroek, Joost and Bussink, Frank and Vidosavljevic, Andrija and Delahaye, Daniel and Aalmoes, Roalt},
  title     = {The influence of traffic structure on airspace capacity},
  year      = {2016},
}

@Article{aarts2023capacity,
  author    = {Aarts, Michiel JM and Ellerbroek, Joost and Knoop, Victor L},
  journal   = {Transportation Research Part C: Emerging Technologies},
  title     = {Capacity of a constrained urban airspace: Influencing factors, analytical modelling and simulations},
  year      = {2023},
  pages     = {104173},
  volume    = {152},
  publisher = {Elsevier},
}

@InProceedings{Hoe:1998,
  author    = {Hoekstra, J and Van Gent, RNHW and Ruigrok, R},
  title     = {Conceptual design of free flight with airborne separation assurance},
  booktitle = {Guidance, Navigation, and Control Conference and Exhibit},
  year      = {1998},
  pages     = {4239},
}

@Article{doole2021constrained,
  author    = {Doole, Malik and Ellerbroek, Joost and Knoop, Victor L and Hoekstra, Jacco M},
  title     = {Constrained Urban Airspace Design for Large-Scale Drone-Based Delivery Traffic},
  journal   = {Aerospace},
  year      = {2021},
  volume    = {8},
  number    = {2},
  pages     = {38},
  publisher = {Multidisciplinary Digital Publishing Institute},
}

@Article{cummings_2022,
  author   = {Christopher Cummings and Hani S. Mahmassani},
  journal  = {Transportation Research Record},
  title    = {Measuring the Impact of Airspace Restrictions on Air Traffic Flow Using Four-Dimensional System Fundamental Diagrams for Urban Air Mobility},
  year     = {2023},
  number   = {1},
  pages    = {1012-1026},
  volume   = {2677},
  doi      = {10.1177/03611981221103237},
}

@Article{cummings_2023,
  author   = {Christopher Cummings and Hani S. Mahmassani},
  journal  = {Transportation Research Record},
  title    = {Comparing Urban Air Mobility Network Airspaces: Experiments and Insights},
  year     = {2023},
  doi      = {10.1177/03611981231185146},
}

@Article{Cum:2021,
  author    = {Cummings, Christopher and Mahmassani, Hani},
  title     = {Emergence of 4-D System Fundamental Diagram in Urban Air Mobility Traffic Flow},
  journal   = {Transportation Research Record},
  year      = {2021},
  volume    = {2675},
  number    = {11},
  pages     = {841--850},
  publisher = {SAGE Publications Sage CA: Los Angeles, CA},
}

@Article{doole2020estimation,
  author    = {Doole, Malik and Ellerbroek, Joost and Hoekstra, Jacco},
  title     = {Estimation of traffic density from drone-based delivery in very low level urban airspace},
  journal   = {Journal of Air Transport Management},
  year      = {2020},
  volume    = {88},
  pages     = {101862},
  publisher = {Elsevier},
}
\end{document}